\RequirePackage{fix-cm}
\documentclass[a4paper, 11pt]{article} 
\usepackage{xcolor}
\usepackage{amsmath}
\usepackage{graphicx,psfrag,epsf}
\usepackage{enumerate}
\usepackage{natbib}
\usepackage{amsfonts}
\usepackage{authblk}

\newcommand{\B}{{\cal B}}
\newcommand{\R}{\mathbb{R}}
\newcommand{\conv}{\mbox{conv}}
\begin{document}

\title{Intraday retail sales forecast: An efficient algorithm for quantile additive modeling} 

\author[1]{M.-O. Boldi\thanks{\texttt{marc-olivier.boldi@unil.ch}}} 
\author[1]{V. Chavez-Demoulin\thanks{\texttt{valerie.chavez@unil.ch}}} 
\author[1]{O. Gallay\thanks{\texttt{olivier.gallay@unil.ch} (corresponding author)}} 

\affil[1]{Faculty of Business and Economics, University of Lausanne, 1015 Lausanne, Switzerland} 








\date{}


\maketitle


\begin{abstract}
\noindent We address the case of a highly frequented retail store where the sales for various grocery products have to be accurately projected. In this context, transforming a stream of Point Of Sale (POS) data into a reliable forecast that evolves over the day is an essential input for any successful replenishment policy to be implemented. To that end, we use quantile regression to adapt different patterns from one product to the other. Dealing with a sustained flow of POS data in real time becomes a challenge as the amount of collected data increases, and  
we develop a stable and efficient quantile additive model algorithm to compute sales forecasts in an intradaily context. We demonstrate the use of our algorithm using POS data from the partner retail store, and it shows to be computationally efficient and therefore suitable for use in real-time dynamic shelf replenishment.   

\vspace{0.5cm}


\end{abstract}




\section{Introduction}
\label{sec:intro}

The present work was motivated by a retail store in the commercial center of the airport of a major European city, which exhibits the highest sales volume in the whole country. The store is open every day of the week from 6am to 11pm. Shopping behavior for groceries can exhibit strong demand variation over the day, and the store needs to be prepared to any high peak of sales at any time and any day of the week. As pointed out by \citep{Mou2018}, the increasing amount and frequency of collected data opens the door for improving sales forecast, both with respect to accuracy and to what can be learnt about the sales patterns. Analyzing and understanding the nature of intraday seasonality in customer demand for fast-moving consumer goods provides opportunities for numerous improvements of in-store processes, ranging from inventory management to workforce scheduling. However, treating in real-time an increased amount of POS data is a challenge. In that regard, we propose an algorithm that makes it easier for companies to translate data that arrives at a higher flow rate into a forecast that evolves in real-time.\\

\begin{figure*}
\centerline{\includegraphics[width=18cm]{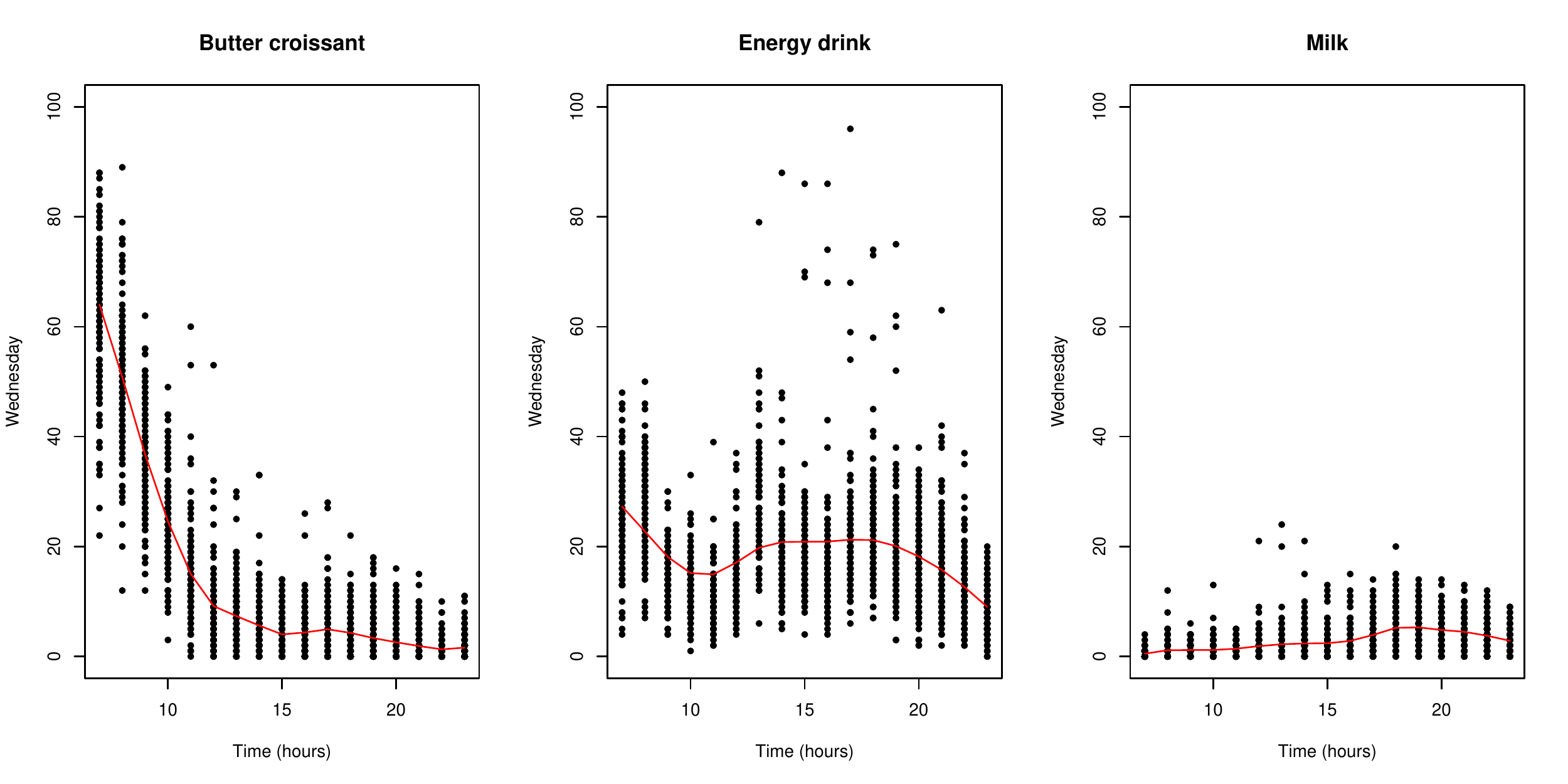}}
\caption{Hourly sales (black points) from 6am to 11pm of butter croissants (left), energy drinks (middle), and milk (right) on Wednesdays from November 1, 2012 to November 23, 2014. The red line shows the mean smoothed by spline.}
\label{data}
\end{figure*}

The considered data consist of hourly sales of various grocery products from November 1, 2012 to November 23, 2014, that is, 743 days times 17 hours. The selected representative products for this study are ``butter croissants'', ``energy drink'', and ``milk'' (one liter). The respective hourly sales on Wednesdays over the observed period are shown in Figure \ref{data}. 
The displayed sales patterns are notably different between the products. Butter croissants are sold mainly early in the morning, with a decreasing trend throughout the day and a small peak of sales around 5pm. Energy drink sales start high, go down around 10am, plateau in the afternoon, and go down again in the evening. Milk sales follow a mild upward trend that dampens in the evening. Note also that the sales are usually higher on Sundays, especially in the afternoon (not shown in Figure \ref{data} but in Figure \ref{Milk}, grey points). This can be explained by the fact that other shops in the area are closed due to city regulations. The sales series considered here are characterized by time-varying mean, variance, and skewness. These features need to be taken into account to help the store understand, model and accurately predict the daily sales pattern. Traditionally, high quantiles are forecasted using quantile regression, 
which can model 
the impact of several covariates on the conditional distribution of the sales 
in a parametric 
way, \citep{ko2005}. The hourly-varying patterns in Figure \ref{data} 
cannot be easily described parametrically. This is why we explore the use of an additive model for quantile regression, also called a quantile additive model \citep{ko2011}, to forecast high quantiles of sales. In the field of statistics, an additive model is a regression method where the effect of the covariates on the response variable does not need to have a parametric form. The idea is to ``let the data speak for themselves" and thanks to the proposed precise representation of when sales occur during the day, it will be possible to predict the moment when specific SKU shelves need to be replenished, and hence to create efficient work schedules. However, fitting a quantile additive model can be challenging because the estimation of such models requires the minimization of an objective function, which can be complex. In particular, in the case of non-differentiability, the classically used gradient descent is unstable. 
Thus, we introduce a simple algorithm combining the gradient sampling algorithm of \citep{BuLeOv2002} and the local scoring algorithm of \citep{HaTi1986}. As a result, the algorithm is stabilized and can be used for non-convex and/or non-differentiable objective functions, as is the case in our additive model framework.
\\

The contribution of this paper is two-fold:  
\begin{itemize}
\item[\textit{(i)}] to introduce the use of a quantile additive model on hourly sales distribution estimates, especially upper quantiles.
\item[\textit{(ii)}] to propose a stable and simple algorithm for the estimation of these models.
\end{itemize}


\vspace{0.2cm}
The aim of the paper is not to make an empirical comparison of the quantile additive model with other existing quantile forecasting methods but to improve its use in a complex context, where frequent and automated updating of forecasts is required for numerous products. Indeed, the performance of quantile additive models has been proven when applied across a variety of different series and compared to competitive alternatives, \citep{horowitz2005, ko2011}. However, when a large number of data series requires an automated procedure, improving the stability and efficiency of the model is crucial.\\ 

The remainder of the paper is organized as follows. In Section \ref{literatureReview}, we give an overview of the general context and the related literature. In Section \ref{quantileAdditive}, we introduce the quantile additive model used in the present case. Section \ref{Estimation} provides details on the model estimation. In Section \ref{gradDescent}, we present the dedicated algorithm, which is introduced for the intraday forecast of retail data. We first introduce the gradient sampling descent algorithm in general, and then combine it with the local scoring algorithm for quantile additive models. In Section \ref{sec_retail_application}, we apply the proposed algorithm to the retail data from the major European retail store. More technical details about the algorithms are described in the Appendix.	




\section{General context and literature review}
\label{literatureReview}

\citep{Mou2018} observe that sales forecasting and its implications for retail inventory management remain of practical importance. 
In particular, sales prediction is highlighted as being fundamental for efficient retail store operations, as it is a key input to the management of stock and personnel. As \citep{Mou2018} states, while backrooms are commonly used in retail stores, there is little discussion on their implementation in the existing literature. However, following \citep{Mou2018}, the study of inventory management processes without distinguishing shelf from backroom inventory is misleading, as lost sales could happen even when the backroom still has stock and customers face empty shelves. This shows the importance of constructing reliable sales forecasts and hence allows for management of the on-shelf availability at an intraday level. Accordingly, in-store processes, which are currently affected by inaccurate forecasting of shopper demand and subsequent replenishment processes, have been identified as one of the major root causes for out-of-stocks, \citep{AaKo2010}. 
While predicting daily total sales in retail operations has been efficient for over a decade \citep{Ar2002}, relatively little consideration has been given to when this total demand actually occurs during the day. Models of in-store shelf replenishment systems have been mainly focused on the replenishment sequences for the SKUs \citep{HaAlMo2007,AbPa2008}, but the question of when to replenish a specific SKU has been rarely addressed. \citep{CuWoFrDoBr2009} and \citep{WoDoBrFr2007} highlight that that sales data should be integrated in shelf replenishment policies and associated inventory control measures to improve operational efficiency. Accordingly, \citep{Le2004} observes that tackling intraday sales patterns is mandatory for an efficient replenishment policy during the different times of the day. In the case of perishable items, on-shelf availability is directly connected to intraday demand patterns in \citep{WoDoBrFr2007}, but no quantification is proposed.\\

As stated in \citep{RaMu2010}, seasonality is acknowledged to be a key element in explaining customer demand patterns. This seasonality in sales may arise at different time granularities: yearly reoccurring variations at particular moments (e.g., Christmas period), specific patterns depending on the day of the week (e.g., highest sales on Saturdays), and non-uniform sales behavior over the day (e.g., more important demand for bread products in the morning). As a result, multiple seasonal patterns must be combined to handle these different levels of granularity \citep{AaKo2010}. \citep{ZoKa2007} highlight that the considered time granularity for the forecast is essential for efficiently predicting sales patterns, but as noted by \citep{ZeDoWoBrFr2009}, little research has been done regarding intraday sales forecasting or for the associated in-store replenishment processes. \citep{Ch2010} claims that setting up a dynamic learning process to analyze sales trends using historical data offers room for retailers to estimate future customer demand more accurately. Some recent contributions (e.g., \citep{ThBu2015}) study how new technologies such as RFID, offering real-time tracking of on-shelf stock level, could replace traditional shelf auditing practices, such as periodic inspections \citep{FiRa2010}, which are currently used by retailers. While these technologies offer more accurate detection of upcoming stock-outs and hence have the potential to both reduce costs and increase service level \citep{ThBu2015}, they lack the predictive dimension that we address in this contribution that opens the door for an efficient management of the workforce over the day. In addition, imperfect detection rate \citep{MeThGeFl2013} and the potentially high set-up costs to monitor all products from the cheapest to the most expensive \citep{PiWoGr2014} are important aspects that are not fully understood yet.\\

Quantile regression is an appealing method for sales prediction because it requires very few assumptions regarding the spread and shape of the sales distribution. The most sophisticated methods based on quantile regression, for example, that of \citep{Ta2007}, generate interval forecasts from quantile predictions using exponentially weighted quantile regression. \citep{ArAh2015} combine a seasonal autoregressive integrated moving average model with quantile regression to construct high and low quantile predictions of daily sales of a perishable product. 

 

\section{An additive model for intraday sales distribution forecast}
\subsection{Additive quantile model for hourly sales}
\label{quantileAdditive}
Let $Y_{tj}$ be the sales at hour $t$ and day $j$, where $t=1,\ldots, T$ and $j=1,\ldots,J$. In the retail data example, we observe $y_{tji}$, $i=1,\ldots,n_{tj}$, with $T=17$, $J=7$. Here, $n_{tj}$ is the number of (assumed) independent replicates of $Y_{tj}$, observed at time $t$ for day $j$.\\ 

In a non-stationary context, $F_{tj}$, the distribution function of $Y_{tj}$, varies with $t$ and $j$. Often, replenishment strategies use the median or more generally a quantile at a pre-defined level $0 < \alpha < 1$, which we write as $q_\alpha(t ; j)$. The level $\alpha$ depends on the target service level of the replenishment strategy. The additive quantile model states that this quantile in day $j$ varies with $t$ as 
\begin{equation}
\label{quantileReg}
q_\alpha(t ; j) = \beta_0 + h_j(t),
\end{equation}
where $h_j$ is some smooth function. This means that the quantile cannot vary too abruptly from one hour to the next. The idea behind smooth $h_j$ is to ``let the data speak for themselves" so that $h_j$ adapts adequately to the $\alpha$-quantile demand pattern on day $j$. A common smoothness requirement is $h_j \in C^2$, meaning that $h_j$ is twice differentiable and that $h_j^{''}$ is continuous. Without further restriction on $h_j$ its location cannot be identified (e.g., if $h_j$ satisfied the smoothness condition, so does $h_j+c$, for any constant $c$), and thus $h_j$ is normalized such that
$$
\int_0^T h_j(t)dt = 0.
$$
So far, with only one covariate, the hour $t$, the model is not additive. However, it can be extended to $d$ covariates, say $x_1,\ldots,x_d$, by writing
\begin{equation*}
q_\alpha(x_1,\ldots,x_d ; j) = \beta_0 + \sum_{k=1}^d h_j(x_k).
\end{equation*}
In particular, the day $j$ can be viewed as a categorical covariate, in which case the additive model is written as
\begin{equation*}
q_\alpha(t ; j) = \beta_j + h(t).
\end{equation*}
This model assumes that the difference between days $j$ and $j'$ is a constant $\beta_{j}-\beta_{j'}$. The within-day variation $h$ is common to all days. A look at the data (see Figures \ref{ButterCroissant}, \ref{Energy}, \ref{Milk}) reveals that this assumption is unreliable, and that one should allow $h$ to vary between days. This variation is called an interaction.


\subsection{Model estimation}
\label{Estimation}
The estimate for day $j$ at time $t$ is obtained by minimizing 
\begin{equation}
\label{functionToMin}
\sum_{t=1}^{T}\sum_{i=1}^{n_{tj}} \rho_{\alpha}\left\{ q_{\alpha}(t;j); y_{tji}\right\},
\end{equation} 
where the risk function $\rho_{\alpha}(q;y)$ is
\begin{equation*}
\rho_\alpha(q;y) = (1-\alpha)(y-q)_- + \alpha (y-q)_+,
\end{equation*}   
$z_+ = \max\{z,0\}$, and $z_-=-\min\{z,0\}$. The smoothness of $h_j$ is obtained by penalizing (\ref{functionToMin}) by the total variation of $h'_j$ 
$$
\lambda \int_0^T |h^{''}_j(t)| dt,
$$
for some $\lambda>0$. A detailed presentation can be found,  for example, in \citep{Koetal1994}. In this paper, we do not explicitly use this penalty in the algorithm that is explained in Section \ref{gradDescent}.


\subsection{The algorithm}
\label{gradDescent}
\subsubsection{Subgradient algorithms}
The subgradient algorithms aim at minimizing \\ $\min_{\mathbf{x} \in D} f(\mathbf{x})$ by iterating for $k=1,2,\ldots$, 
\begin{equation}
\label{StepEq}
\mathbf{x}^{(k+1)} = \mathbf{x}^{(k)} - s_k\mathbf{d}^{(k)},
\end{equation} 
The step $\mathbf{d}^{(k)}$ is a subgradient of $f$ at $\mathbf{x}^{(k)}$: 
\begin{eqnarray*}
\mathbf{d}^{(k)} &\in& \partial f(\mathbf{x}^{(k)}) \\
&&= \left\{\mathbf{d} : f(\mathbf{x}) \geq f(\mathbf{x}^{(k)}) + \mathbf{d}^T (\mathbf{x}-\mathbf{x}^{(k)}), \forall \mathbf{x}) \right\}.
\end{eqnarray*}
When $f$ is differentiable at $\mathbf{x}$, $\partial f(\mathbf{x})$ reduces to $\{\nabla f(\mathbf{x})\}$. The step sizes $s_k>0$ can be selected using the backtracking line search: scan $s \in \{1,1/2, 1/4, \ldots\}$ until the Armijo's condition below is satisfied:
\begin{equation}
f(\mathbf{x}^{(k)} + s\mathbf{d}^{(k)}) < f(\mathbf{x}^{(k)})- s \mathbf{d}^T\mathbf{d}.
\label{linesearch}
\end{equation} 

When $f$ is differentiable at $\mathbf{x}^{(k)}$, the algorithm step is a gradient descent, $\mathbf{d}_k$ being $\{\nabla f(\mathbf{x})\}$. When $f$ is not differentiable at $\mathbf{x}^{(k)}$, the algorithm is not guaranteed to be a descent algorithm. In this case, one needs to track the achieved minimum in the past steps. Furthermore, the subgradient may not be computable exactly except in specific cases. \citep{BuLeOv2002} proposed a gradient sampling algorithm to approximate the Clarke subgradient, a good candidate in $\partial f(\mathbf{x}^{(k)})$, which we now use.  


\subsubsection{Gradient sampling algorithm}
\label{GSA}
The gradient sampling algorithm (GSA) was first introduced by \citep{BuLeOv2002} and has been further developed by \citep{BuLeOv2005}, \citep{Ki2007}, and \citep{Ki2010}. An up-to-date presentation can be found in Chapter 13 of \citep{BaKaMa2014}.\\ 

For the sake of completeness, below we quote some definitions following \citep{Ki2007}, adapted to our context. Let $f:\R^n\rightarrow \R$ be a locally Lipschitz continuous function, continuously differentiable\footnote{In the Lebesgue sense, that is the set of points where $f$ is non-continuously differentiable is of null Lebesgue measure.} on an open dense set $D\in \R^n$. The Clarke subdifferential of $f$ at $\mathbf{x}\in \R^n$ is the set
\begin{equation*}
\bar{\partial}f(\mathbf{x}) = \mbox{conv}\left\{\lim_j \nabla f(\mathbf{z}^j):\mathbf{z}^j\rightarrow \mathbf{x}, \; \mathbf{z}^j \in D\right\},
\end{equation*}
where conv$(A)$ is the convex hull of $A$. A point $\mathbf{x}$ is called stationary for $f$ if $\mathbf{0} \in \bar{\partial}f(\mathbf{x})$, where $\mathbf{0}$ is the $n$-vector of 0's. The Clarke $\epsilon$-subdifferential is defined by $\bar{\partial}_\epsilon f(\mathbf{x}) = \mbox{conv}\left[\bar{\partial}f\{B(\mathbf{x},\epsilon)\}\right]$, where $B(\mathbf{x},\epsilon)$ is a ball of radius $\epsilon \geq 0$ centered at $\mathbf{x}$. A point $\mathbf{x}$ is called $\epsilon$-stationary for $f$ if $\mathbf{0} \in \bar{\partial}_\epsilon f(\mathbf{x})$. For a closed convex set $G$, let $\mbox{Proj}(\mathbf{0}|G)$ the projection of $\mathbf{0}$ onto $G$. The Clarke subgradient of $f$ at $\mathbf{x}$ is defined by $g_\epsilon(\mathbf{x})=\mbox{Proj}\{\mathbf{0}|\bar{\partial}_\epsilon f(\mathbf{x})\}$. Denoted by cl conv$(A)$, the closed convex hull of $A$, the Clarke $\epsilon$-subdifferential $\bar{\partial}_\epsilon f(\mathbf{x})$ is approximated by
\begin{equation*}
G_\epsilon (\mathbf{x}) = \mbox{cl conv}\left[\nabla f\{B(\mathbf{x},\epsilon)\cap D\}\right],
\end{equation*}
since $G_\epsilon(\mathbf{x})\subset \bar{\partial}_\epsilon f(\mathbf{x})$ and $\bar{\partial}_{\epsilon_1} f(\mathbf{x})\subset G_{\epsilon_2}(\mathbf{x})$ for any $0\leq \epsilon_1 < \epsilon_2$. To put that in action, the GSA simulates independent $\{\mathbf{x}_i\}_{i=1}^m$ in $\B(\mathbf{x},\epsilon)$ to approximate $G_\epsilon(\mathbf{x})$ by 
\begin{equation*}
\hat{G}_\epsilon(\mathbf{x}) = \mbox{conv} \{\nabla f(\mathbf{x}),\nabla f(\mathbf{x}_1),\ldots,\nabla f(\mathbf{x}_m)\},
\end{equation*}
where $m \geq n+1$. Next, $g_\epsilon(\mathbf{x})$ is approximated by $\Hat{g}_\epsilon(\mathbf{x})=\mbox{Proj}\{\mathbf{0}|\Hat{G}_{\epsilon}(\mathbf{x})\}$. Finally, $-\hat{g}_\epsilon(\mathbf{x})$ is used as the step $\mathbf{d}$ in (\ref{StepEq}). 
The full GSA algorithm is reported in Appendix~\ref{Appendix}. The main step is the computation of $\hat{g}_\epsilon(\mathbf{x})$, as follows
\begin{enumerate}
\item Draw independent $\mathbf{u}_1, \ldots, \mathbf{u}_m$ on the unit ball ${\cal B}(\mathbf{0},1)$.
\item Set $\hat{G}_\epsilon(\mathbf{x}) = \{\nabla f(\mathbf{x}), \nabla f(\mathbf{x} + \epsilon \mathbf{u}_1), \ldots, \nabla f(\mathbf{x} + \epsilon \mathbf{u}_m)\}$.
\item Find 
\begin{equation} \hat{g}_\epsilon(\mathbf{x}) = \arg \min\left[\Vert \mathbf{g} \Vert: \mathbf{g} \in \conv \left\{\hat{G}_\epsilon(\mathbf{x})\right\} \right].
\label{Clark}
\end{equation} 
\end{enumerate}
Subproblem (\ref{Clark}) happens to be quadratic and fairly easy to solve. These details are reported in Appendix~\ref{Appendix}. This GSA step is thus quite easy to implement in practice and provides a general way to perform subgradient-based optimization.\\

\begin{figure*}
\centerline{\includegraphics[width=14cm]{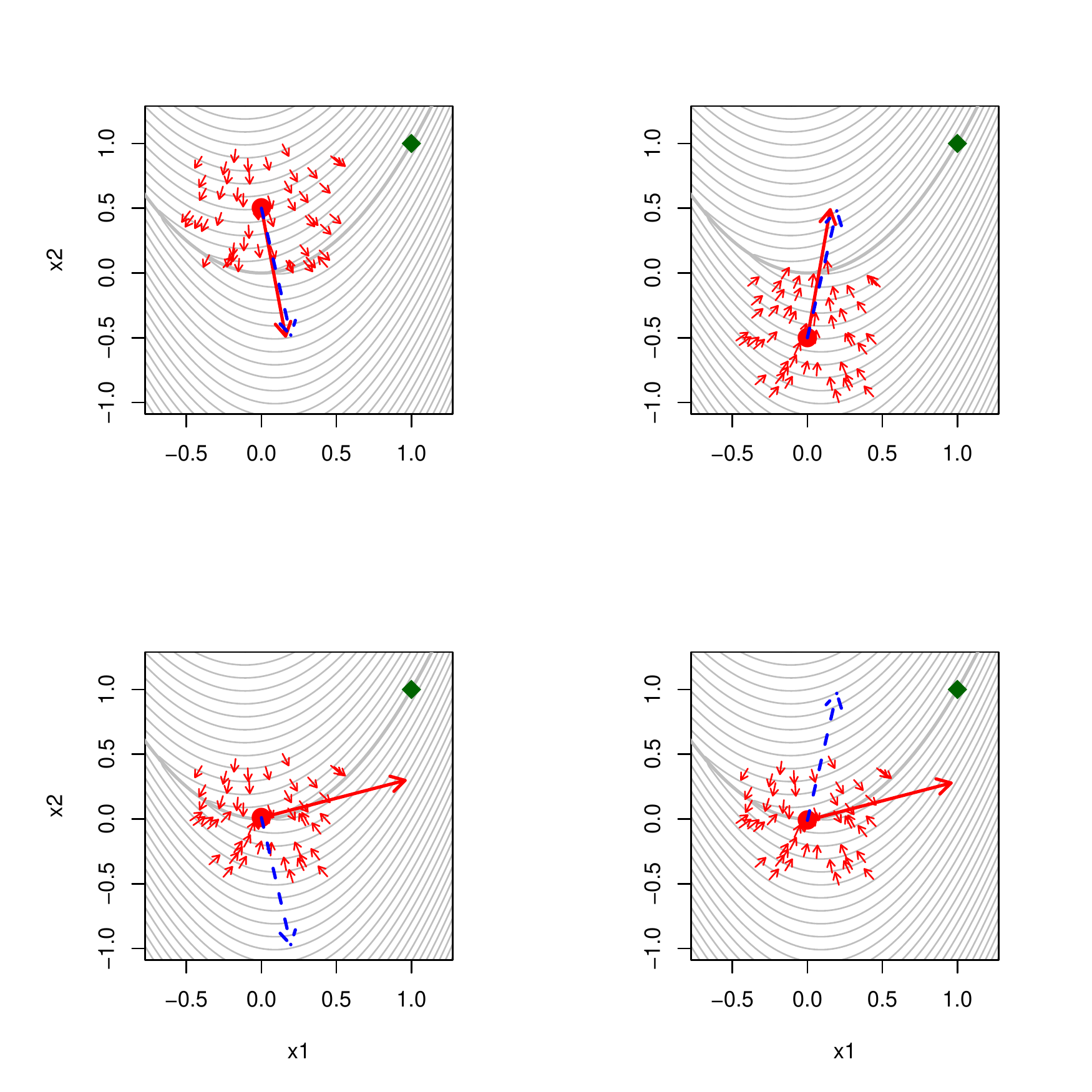}}
\caption{Four cases with function $f(\mathbf{x})=10(x_2-x_1^2)+(1-x_1)^2$. The diamond is at the minimum $(1,1)^T$. The big dot is the current solution $\mathbf{x}$. The bold line is the non-differentiability set $\Omega=\{x_2=x_1^2\}$. The dashed arrow is the gradient descent $-\nabla f(\mathbf{x})$. The short arrows are the sampled gradient descents $-\nabla f(\mathbf{x} + \epsilon \mathbf{u}_1), \ldots, -\nabla f(\mathbf{x} + \epsilon \mathbf{u}_m)$. The solid bold arrow is the approximation descent $\hat{g}_\epsilon(\mathbf{x})$. The lengths of all the arrows have been modified to facilitate readability. Top left plot: current solution $\mathbf{x}$ is far above $\Omega$. Top right plot: $\mathbf{x}$ is far below $\Omega$. Bottom left plot: $\mathbf{x}$ is close above $\Omega$. Bottom right plot: $\mathbf{x}$ is close below $\Omega$.}
\label{GS_illustration}
\end{figure*}

Borrowed from \citep{Ov2015}, the GSA is illustrated in Figure~\ref{GS_illustration}. The four plots show cases where the current solution $\mathbf{x}$ is either far from or close to the non-differentiable area. In the two top plots, $f$ is differentiable at $\mathbf{x}$, and the GSA step is like a gradient descent step. In the two bottom plots, the GSA step is robust and points in the good direction while the gradient descent alternates up (right plot) and down (left plot), suffering numerical instability due to the non-differentiability. Thus, in practice, using the GSA avoids the typical zigzag behavior of gradient algorithms. However, using the GSA comes at the cost of several gradient computations, making it inappropriate in large dimensions or complex cases without further adaptations.\\ 

\vspace{-0.3cm} 
In Section \ref{quantile}, we combine the GSA with a local scoring algorithm for the case of quantile additive models.  
\subsection{GSA algorithm for quantile additive models}
\label{quantile}
The GSA can be used to solve (\ref{functionToMin}) using the derivative of $\rho$ at $q\neq y$:
\begin{equation}
\nabla \rho_\alpha(q;y) = \left\{
\begin{array}{rl}
1 - \alpha, & \mbox{ if } y-q < 0,\\
- \alpha, & \mbox{ if } y-q > 0.
\end{array}
\right.
\label{diff}
\end{equation}
The application would be straightforward if $q$ is assumed to be a linear function $\mathbf{w}^T\beta$ of some covariates $\mathbf{w}$, or any other parametric form. In our application, we want a non-parametric smooth solution. One possibility would be to add a penalty to the objective function (\ref{functionToMin}). Instead, we used a local scoring algorithm (LSA).\\

The local scoring algorithm was first introduced by \citep{HaTi1986}. It can be seen as a special case of a projected gradient algorithm such that the final solution has some predefined properties, for example, smoothness and additivity in covariates. To achieve this, a scatter plot smoother $h$ is applied to the direction $\mathbf{d}$ at each step of the algorithm:
\begin{eqnarray*}
\nonumber \mathbf{x}^{(k+1)} &=& \mathbf{x}^{(k)} + s h(\mathbf{d}^{(k)}), 
\mbox{\quad \quad\quad or} \mathbf{x}^{(k+1)} = h\left(\mathbf{x}^{(k)} + s h(\mathbf{d}^{(k)})\right).
\end{eqnarray*}
\citep{HaTi1986} leave open the discussion about the convexity of the projection. They allow the possibility to smooth both $\mathbf{d}^{(k)}$ and $\mathbf{x}^{(k+1)}$ to guaranty that at each step the new proposal is smooth. In the application, $h_j$ is the LOESS smoother (see e.g., \citep{ChHa1992}) with the covariate $t$. It can be conveniently obtained using the {\tt R} package {\tt mcgv} and its function {\tt gam} (see \citep{Wo2006}).\\ 

Combining the GSA and the LSA gives the following algorithm main step from $k$ to $k+1$:
\begin{enumerate}
\item Draw independent $\mathbf{u}_1, \ldots, \mathbf{u}_m$ on the unit ball ${\cal B}(\mathbf{0},1)$.
\item Set $\hat{G}_\epsilon^{(k)}$ equal to
$$
\left\{\nabla \rho_\alpha(\mathbf{q}^{(k)}),\nabla \rho_\alpha(\mathbf{q}^{(k)} + \epsilon \mathbf{u}_1),\ldots,\nabla \rho_\alpha(\mathbf{q}^{(k)} + \epsilon \mathbf{u}_m)\right\}
$$
\item Set $\mathbf{g}^{(k+1)}$ equal to $\arg \min\left\{\Vert \mathbf{g} \Vert: \mathbf{g} \in \conv\, \hat{G}_\epsilon^{(k)} \right\}$.
\item Set $\mathbf{d}^{(k+1)} = -h(\mathbf{g}^{(k+1)})$.
\item Update $\mathbf{q}^{(k+1)} = \mathbf{q}^{(k)}+ s \mathbf{d}^{(k+1)}$, where $s$ is appropriately selected.
\end{enumerate}
The initial solution $\mathbf{q}^{(0)}$ can be a constant quantile or a more sophisticated estimate. More details about the algorithm are provided in Appendix~\ref{Appendix}, especially in terms of the selection of $s$, and an approximate solution of (\ref{Clark}).

\section{Application of GSA to retail data}
\label{sec_retail_application}
\begin{figure*}
\centerline{\includegraphics[width=15cm]{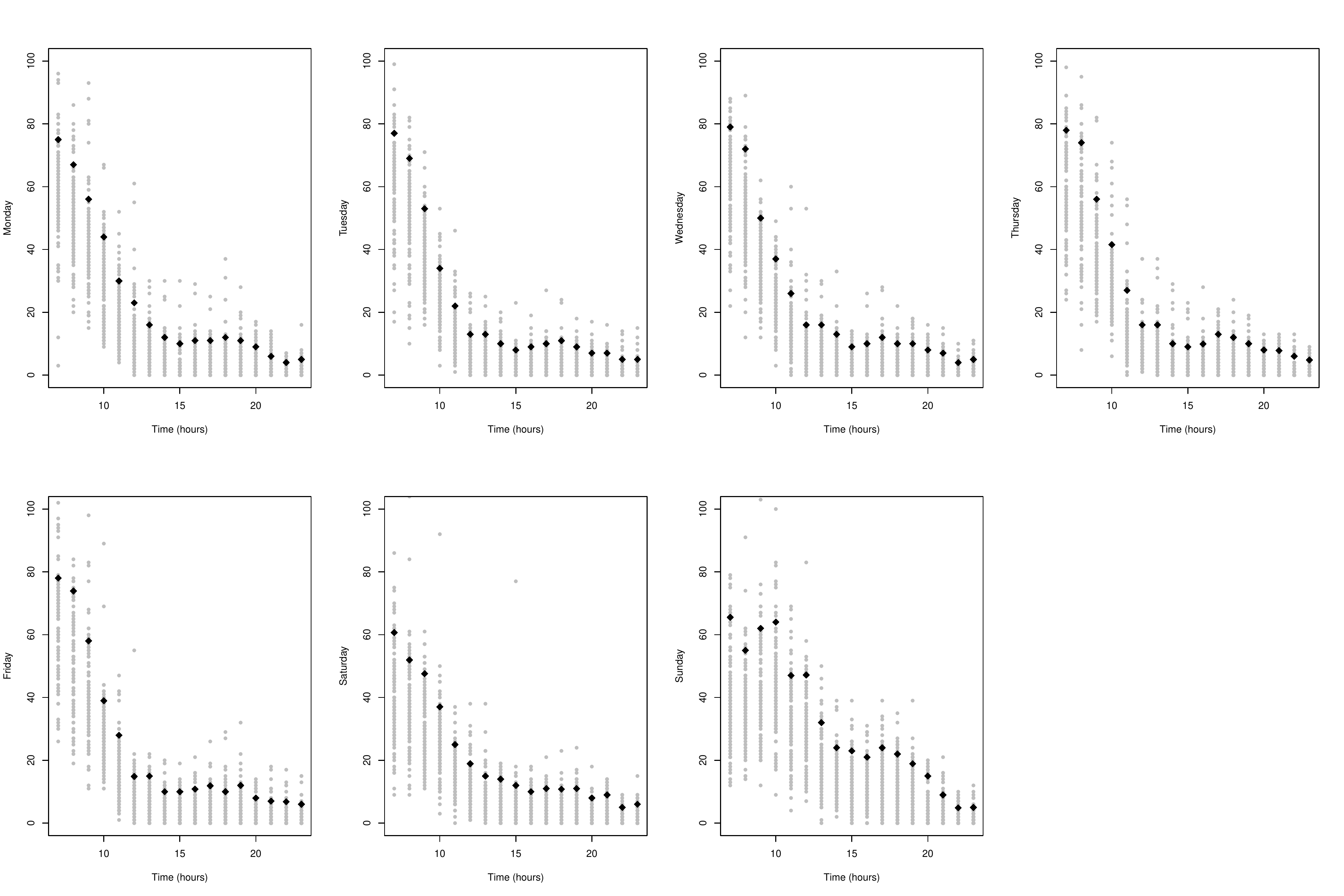}}
\caption{Intradaily sales (points in grey) of butter croissants at the seven days of the week (panels) from November 1, 2012 to November 23, 2014. The black points are the 90\%-quantile estimates.}
\label{ButterCroissant}
\end{figure*}

\begin{figure*}
\centerline{\includegraphics[width=15cm]{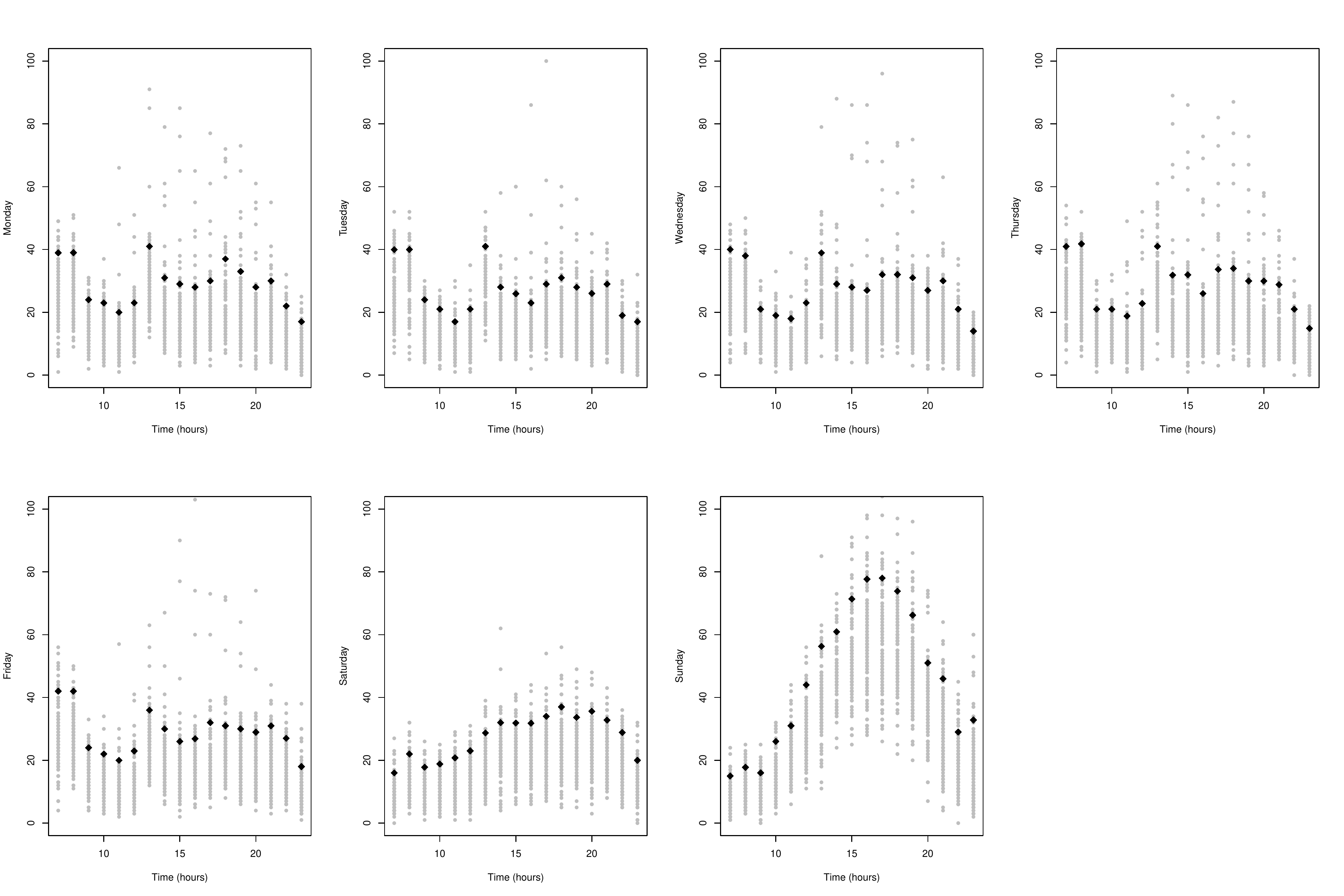}}
\caption{Intradaily sales (points in grey) of energy drink at the seven days of the week (panels) from November 1, 2012 to November 23, 2014. The black points are the 90\%-quantile estimates.}
\label{Energy}
\end{figure*}

\begin{figure*}
\centerline{\includegraphics[width=15cm]{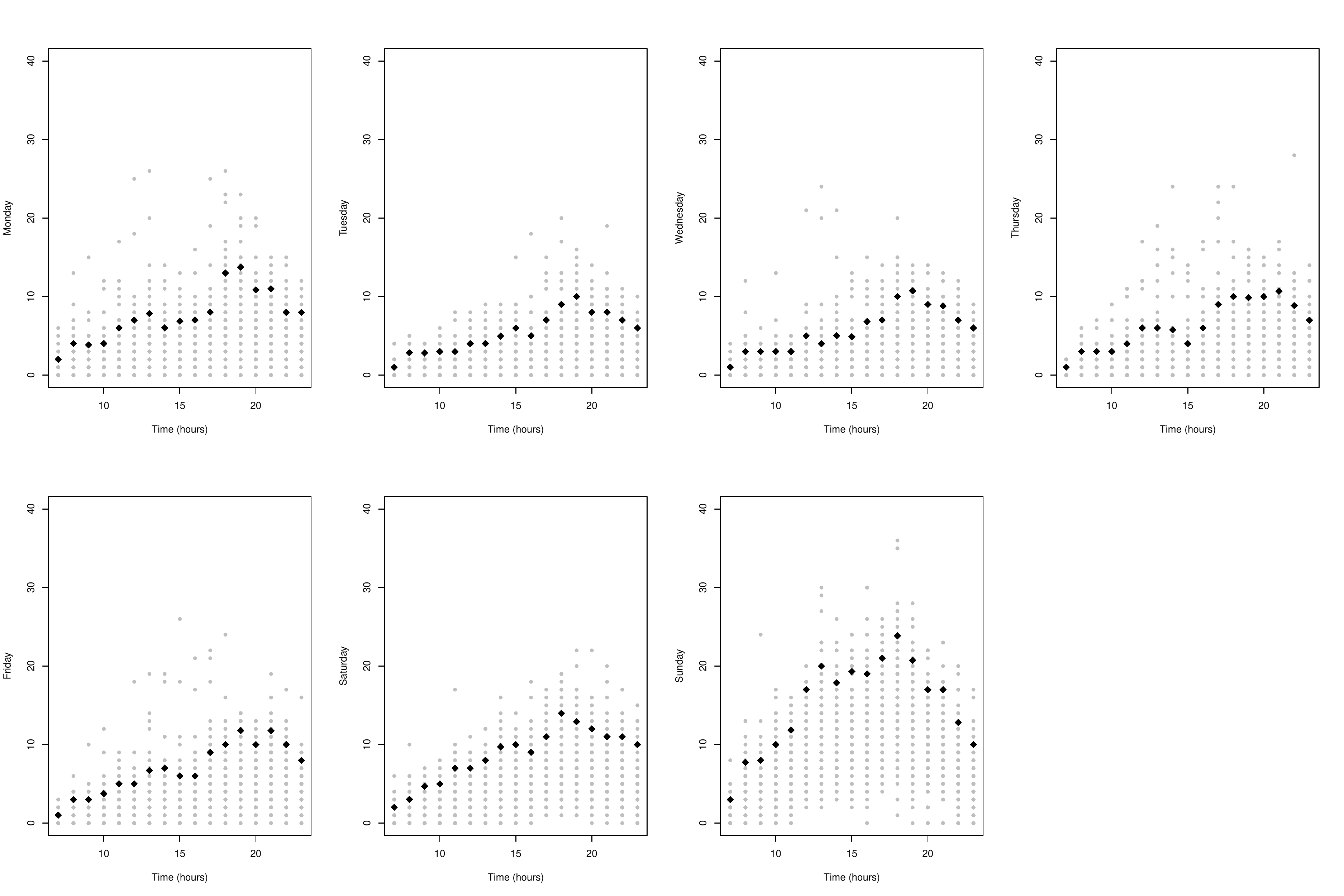}}
\caption{Intradaily sales (points in grey) of milk (one liter) for each of the seven days of the week (panels) from November 1, 2012 to November 23, 2014. The black points are the 90\%-quantile estimates. }
\label{Milk}
\end{figure*}

We consider the European retail store with three selected product sales series represented in Figure \ref{data} and in Figures \ref{ButterCroissant}, \ref{Energy}, and \ref{Milk} (grey points), which are characterized by hourly-varying mean, variance, and skewness. Traditional forecasting methods fail to capture these features for two reasons: first, the commonly used methods, such as the exponentially weighted moving average model, generalized linear model, or generalized additive model, are modelling the mean sales and therefore are not robust to outliers. That is, one exceptional sale may influence the entire mean estimation. Second, the commonly used parametric models require an assumption to be made about the sales distribution, such as the Poisson distribution. With the observed features of our dataset, it is preferable to avoid any assumption about the sales and to adapt a nonparametric model. Beside the robustness to outliers and nonparametric properties of the quantile additive model, another advantage compared to any empirical method (e.g., calculating the quantile at each hour of each day) is the formal inferential framework it offers, \citep{ko2011}. As for any regression type model, all the hourly sales series are pooled together over all days so that the quantiles are estimated by the significance of the covariates in relation to each other. Note that with the purpose of shelf-replenishment planning over all the products of the store, the quantile model may be applied to a dataset where all the store products series are pooled together. The quantile additive model is a well-established method used in many fields of applied statistics and the aim of our application is hence not to compare it with other forecasting methods but rather to illustrate the use of our efficient algorithm when the quantile additive model is applied to complex real data and to interpret the usefulness of the results.\\ 

Going back to the retail data, the store manager is given the freedom to choose a target service level, depending on the business context he or she is specifically facing. This service level makes it possible to determine what guarantees the store wants to allocate regarding the availability of a product. For the various and heterogeneous products introduced in Section \ref{sec:intro}, we choose a service level of $\alpha=90\%$ in model (\ref{quantileReg}). Following this, in order to ensure that the product shelf is never empty at any time of the day, a quantity of interest would be the 90\% quantile of sales. As expected, the algorithm quickly converges. More precisely, when executed on a personal laptop (2.60 GHz Intel i7-6500U processor with 12 GB RAM available), for a matrix of 753$\times$18 data, the CPU time required for execution by the system on behalf of the calling process is less than 9 seconds, and the CPU time required for the execution of user instructions of the calling process is around 4 minutes. The resulting estimated quantiles are shown in Figure \ref{ButterCroissant} for the butter croissants,  Figure \ref{Energy} for the energy drink, and  Figure \ref{Milk} for the milk (black points). 
The interaction between day of the week and hour of the day allows the 
90\%--quantile estimate to capture simultaneously the intradaily and intraweek patterns.\\

The generated extra information on the sales patterns over the day should be taken as an input by any implemented replenishment policy, as significant improvements would be expected. 
Indeed, this increased knowledge about the sales fluctuation during the day allows for finer-grained shelf replenishment strategies that would decrease the probability of stock-out. Furthermore, it would open the door for better management of personal scheduling. Typically, we can observe in Figures \ref{ButterCroissant} and \ref{Energy} that butter croissants should be replenished earlier during the day compared to the energy drink. The proposed methodology not only allows for a qualitative and visual analysis of this assertion but also offers a framework that allows for a complete quantitative and automatic treatment of these aspects.
\section{Conclusion}
\label{conclusion}
In this paper, directly motivated by the need for precise intraday forecast for improving shelf replenishment operations, we adapt state-of-the-art algorithms and apply them to this particular context. More precisely, the standard gradient sampling algorithm
is adapted by incorporating the local scoring algorithm. The proposed methodology helps to solve the complex estimation problems that typically arise in the context of intraday sales forecasting for retail operations, specifically to overcome non-differentiability and/or non-smoothness when optimizing the resulting objective function (a log-likelihood and a risk function) under some constraints of additivity and smoothness of the parameters. 
The algorithm is very flexible and easy to implement, and its computational performance allows for  wide implementation in real-time frameworks, and hence it generally offers an efficient alternative to the Frisch--Newton interior point method of \citep{PoKo1997}.
\bibliographystyle{spbasic}      
\bibliography{Methodology}

%
%


\appendix
\section{Appendix}
\label{Appendix}
\subsection{Gradient sampling descent algorithm}
\label{GSDA:app}
The gradient sampling algorithm of \citep{Ki2007} is reported below, borrowed from \citep{Ov2015}.
\begin{enumerate}
\item Fix the sampling size $m \geq n+1$, a line search parameter $0 < \beta < 1$, and the reduction factors $\mu\in (0,1)$ and $\lambda \in (0,1)$.
\item Initialize the solution $\mathbf{x}$, the radius $\varepsilon > 0$ and the tolerance $\tau > 0$.
\item Compute the approximation $\hat{g}_\varepsilon(\mathbf{x})$.
\begin{enumerate}
\item Sample $\mathbf{u}_1, \ldots, \mathbf{u}_m$ on the unit ball ${\cal B}(\mathbf{0},1)$, where $\mathbf{0}$ is the $n$-vector of 0.
\item Set $\hat{G}_\varepsilon(\mathbf{x}) = \{\nabla f(\mathbf{x}), \nabla f(\mathbf{x} + \varepsilon \mathbf{u}_1), \ldots, \nabla f(\mathbf{x} + \varepsilon \mathbf{u}_m)\}.$
\item Find 
\begin{equation} \hat{g}_\varepsilon(\mathbf{x}) = \arg \min\left[\Vert g \Vert: g \in \conv \left\{\hat{G}_\varepsilon(\mathbf{x})\right\} \right].
\label{quadprod:app}
\end{equation} 
\end{enumerate}
\item If $\Vert \hat{g}_\varepsilon(\mathbf{x}) \Vert \leq \tau$, update $\varepsilon \leftarrow \mu \varepsilon$ and $\tau \leftarrow \lambda \tau$, go to 3.
\item If $\Vert \hat{g}_\varepsilon(\mathbf{x}) \Vert > \tau$, do a backtracking line search along $\mathbf{d}=-g_\varepsilon(\mathbf{x})/\Vert g_\varepsilon(\mathbf{x}) \Vert$, diminishing $t \in \{1, 1/2, 1/4, \ldots\}$ until the Armijo's condition is satisfied,
\begin{equation}
f(\mathbf{x} + t\mathbf{d}) < f(\mathbf{x})-\beta t \mathbf{d}^T\nabla f(\mathbf{x}).
\label{linesearch}
\end{equation}
\item Update $\mathbf{x} \leftarrow \mathbf{x}+t\mathbf{d}$ and go to 3.
\end{enumerate}
The overall algorithm is stopped when one or several convergence criteria are satisfied, typically, when $\tau$ and $\varepsilon$ reach a predefined threshold.

To solve sub-problem~(\ref{quadprod:app}), recall that any element of $\conv(G)$ can be written as a unique convex combination of vectors at the edges of $\conv(G)$, say $G_e=\{\mathbf{z}_1, \ldots ,\mathbf{z}_m\}$. In more detail, let $\mathbf{z} \in \conv(G)$, and then there exists an unique $\mathbf{r}$ such that:
\begin{equation*}
\mathbf{z} = r_1 \mathbf{z}_1 + \dots + r_m\mathbf{z}_l  = Z\mathbf{r}, \quad \mathbf{r} \succeq 0, \quad \sum_{j=1}^m r_j = 1,
\end{equation*}
where $Z$ is the matrix with columns $\{\mathbf{z}_1, \ldots, \mathbf{z}_l\}$. \citep{Ed1977} shows how to derive the columns of $Z$. The method is efficiently implemented in {\tt chull} of {\tt R}. Because there exits $\mathbf{r}^*$ such that $g=Z\mathbf{r}^*$ for each $g\in \conv \left\{\hat{G}_\varepsilon(\mathbf{x})\right\}$, the sub-problem~(\ref{quadprod:app}) can be written as the quadratic problem under linear constraints:
\begin{eqnarray*}
\min_{\mathbf{r}  \in \R^m} && \mathbf{r} ^T Z^TZ \mathbf{r}\\
s.t. && \mathbf{r} \succeq 0,\\
&& \sum_{j=1}^m r_j = 1.
\end{eqnarray*}
This minimization problem can be efficiently solved using classical quadratic programming (e.g., function {\tt solveQP} of {\tt R} package {\tt quadprog}; \citep{quadprog}). To further simplify this step, we have noted that, at certain iterations of the algorithm, the above problem may be numerically unstable or difficult to compute. In such case, $
\hat{g}_\varepsilon(\mathbf{x})$ can be conveniently replaced by the average $(m+1)^{-1} \{\nabla f(\mathbf{x})+\sum_{j=1}^m\nabla f(\mathbf{x}+\varepsilon \mathbf{u}_j)\}$. Although we have not found a formal proof, this is intuitive since any stable vector pointing toward $\hat{G}_\varepsilon(\mathbf{x})$ could be used as an approximation of $g_\varepsilon(\mathbf{x})$.

Finally, the Armijo's conditions $(\ref{linesearch})$ can be replaced by $f(\mathbf{x} + t\mathbf{d}) < f(\mathbf{x})-\beta t \Vert \hat{g}_\varepsilon(\mathbf{x}) \Vert$. Indeed, as an approximation  of $\nabla f(\mathbf{x})$, $\hat{g}_\varepsilon(\mathbf{x})$ can be used in the right-hand side of $(\ref{linesearch})$, giving $\mathbf{d}^T\hat{g}_\varepsilon(\mathbf{x}) =  \Vert\hat{g}_\varepsilon(\mathbf{x})\Vert$, see, for instance, \citep{Ov2015}.
\subsection{GSDA for quantile additive models}
The complete algorithm when applied to quantile additive models is given below. 
\begin{enumerate}
\item Fix the sampling size $m \geq n+1$, a line search parameter $0 < \beta < 1$, and the reduction factors $\mu\in (0,1)$ and $\lambda \in (0,1)$.
\item Initialize the radius $\varepsilon > 0$ and the tolerance $\tau > 0$.
\item Initialize the solution $\mathbf{q}_\alpha$.
\item Compute the approximation $\hat{g}_\varepsilon(\mathbf{q}_\alpha)$.
\begin{enumerate}
\item Sample $\mathbf{u}_1, \ldots, \mathbf{u}_m$ on the unit ball ${\cal B}(\mathbf{0},1)$, where $\mathbf{0}$ is the $n$-vector of 0.
\item Set 
\begin{equation*}
\hat{g}_\varepsilon(\mathbf{q}_\alpha) = (m+1)^{-1}\left\{\nabla \rho_\alpha(\mathbf{q}_\alpha)+\sum_{k=1}^m \nabla \rho_\alpha(\mathbf{q}_\alpha + \varepsilon \mathbf{u}_k)\right\}
\end{equation*}
\end{enumerate}
\item If $\Vert \hat{g}_\varepsilon(\mathbf{q}_\alpha) \Vert \leq \tau$, update $\varepsilon \leftarrow \mu \varepsilon$ and $\tau \leftarrow \lambda \tau$, go to 4.
\item If $\Vert \hat{g}_\varepsilon(\mathbf{q}_\alpha) \Vert > \tau$.
\begin{enumerate}
\item Set $\mathbf{d}^*= -{\tt gam}(\hat{g}_\varepsilon(\mathbf{q}_\alpha) \sim {\tt lo}(s ,by=j)$
and $\mathbf{d}=\mathbf{d}^*/\Vert \mathbf{d}^*\Vert$.
\item Diminish $t \in \{1, 1/2, 1/4, \ldots\}$ until $
\rho_\alpha(\mathbf{q}_\alpha + t\mathbf{d}) < \rho_\alpha(\mathbf{q}_\alpha)-\beta t \mathbf{d}^T\nabla \rho_\alpha(\mathbf{q}_\alpha)$.
\end{enumerate}
\item Update $\mathbf{q}_\alpha \leftarrow \mathbf{q}_\alpha+t\mathbf{d}$ and go to 4.
\end{enumerate}
As in Section~\ref{GSDA:app}, the algorithm can be adapted, for example, replacing the average in step~4(b) with the solution to sub-problem~(\ref{quadprod:app}).

\end{document}